\documentclass[10pt, conference, letterpaper]{IEEEtran}

\IEEEoverridecommandlockouts

\usepackage{cite}
\usepackage{subfig}
\usepackage[colorinlistoftodos]{todonotes}
\usepackage{amsmath,amssymb,amsfonts}
\usepackage{algorithmic}
\usepackage{graphicx}
\usepackage{textcomp}
\usepackage{xcolor}

\def\BibTeX{{\rm B\kern-.05em{\sc i\kern-.025em b}\kern-.08em
    T\kern-.1667em\lower.7ex\hbox{E}\kern-.125emX}}
\begin{document}

\graphicspath{{figs/}}
\hyphenation{}

\newcommand{\etal}    {\textit{et al.}}
\newcommand{\tw}[1]   {\textcolor{blue}{\textbf{[Thomas] #1}}}
\newcommand{\db}[1]   {\textcolor{blue}{\textbf{[David] #1}}}
\newcommand{\is}[1] 
{\textcolor{red}{\textbf{[Ioana] #1}}}
\title{Experimental Clock Calibration\\on a Crystal-Free Mote-on-a-Chip}

\author{
    \IEEEauthorblockN{
        Ioana Suciu$^1$,
        Filip Maksimovic$^2$,
        David Burnett$^2$,
        Osama Khan$^2$,
        Brad Wheeler$^2$, \\
        Arvind Sundararajan$^2$,
        Thomas Watteyne$^3$,
        Xavier Vilajosana$^1$\,$^4$,
        Kris Pister$^2$
    }
    \IEEEauthorblockA{
        $^1$~Worldsensing, Spain. {\tt isuciu@worldsensing.com}\\
        $^2$~University of California, Berkeley, USA. {\tt \{fil,db,oukhan,brad.wheeler,arvinds,ksjp\}@berkeley.edu}\\
        $^3$~Inria, France. {\tt thomas.watteyne@inria.fr}\\
        $^4$~Universitat Oberta de Catalunya and Worldsensing, Spain {\tt xvilajosana@uoc.edu}
    }
}

\maketitle

\begin{abstract}
The elimination of the off-chip frequency reference, typically a crystal oscillator, would bring important benefits in terms of size, price and energy efficiency to IEEE802.15.4 compliant radios and systems-on-chip.
The stability of on-chip oscillators is orders of magnitude worse than that of a crystal.
It is known that as the temperature changes, they can drift more than $50 ppm/^{\circ} C$.
This paper presents the result of an extensive experimental study. 
First, we propose mechanisms for crystal-free radios to be able to
    track an IEEE802.15.4 join proxy,
    calibrate the on-chip oscillators and
    maintain calibration against temperature changes.
Then, we implement the resulting algorithms on a crystal-free platform and present the results of an experimental validation.
We show that our approach is able to track a crystal-based IEEE802.15.4-compliant join proxy and maintain the requested radio frequency stability of $\pm 40 ppm$, even when subject to temperature variation of $2^{\circ} C/min$.
\end{abstract}

\begin{IEEEkeywords}
    crystal-free radio,
    IEEE802.15.4,
    short-range wireless,
    standard-compliance,
    clock calibration,
    reference frequency stability,
    low-power wireless mesh networking.
\end{IEEEkeywords}

\section{Introduction}


IEEE802.15.4~\cite{IEEE802154} is a popular standard for short-range wireless communication.
IEEE802.15.4-compliant radios typically operate in the 2.4~GHz frequency band.
It is the underlying technology in protocol stacks such as ZigBee or 6TiSCH, which organize a number of IEEE802.15.4-compliant devices in a mesh network topology.
These types of networks are widely used in home automation, smart building, smart city and industrial applications.


Manufacturers building IEEE802.15.4-compatible chips must ensure a radio frequency stability not exceeding $\pm 40 ppm$ while transmitting a packet~\cite{NXPreferenceXTAL}.
Crystals are typically used as the reference oscillators for synthesizing the radio frequency, and keeping time.
Crystals are popular because they provide the necessary stability while remaining low-power.
It is common to use two crystals:
    a ``slow'' ultra low-power crystal oscillator for time-keeping (typ. 32,768~Hz) and
    a ``fast'' and more power-hungry crystal oscillator for radio frequency synthesis (typ. 12-48~MHz)~\cite{griffith17referenceXTALless}.
Crystal oscillators are, however, off-chip elements.
They contribute to the energy consumption, size and cost of the final product, which becomes significant at high volumes (millions of chips)~\cite{khan18time}. 


To further integrate IEEE802.15.4-compliant chips, research is being done on designing on-chip oscillators that could replace the off-chip (crystal) oscillators~\cite{shih13tunableOnchip,zhang11selfCal,Burn1901:CMOS}, aiming to obtain crystal-free architectures.
Not needing to add crystals to a radio has several key advantages.
All elements -- including the oscillators -- are part of a single piece of silicon.
This significantly reduces price of the device.
It also reduces its footprint (size) in a design, to the point that it can be used completely standalone, provided it embeds a power source.
In the extreme case, such a mote-on-a-chip can even be considered disposable.
There are some further benefits, such as the fact that the start-up time of an on-chip resonator is much 
shorter than that of a crystal, making the system faster to switch on and off.
The resulting more efficient duty-cycling further reduces the average power consumption of the crystal-free modules \cite{khan18time}.

\begin{figure}
    \centering
    \includegraphics[width=0.85\columnwidth]{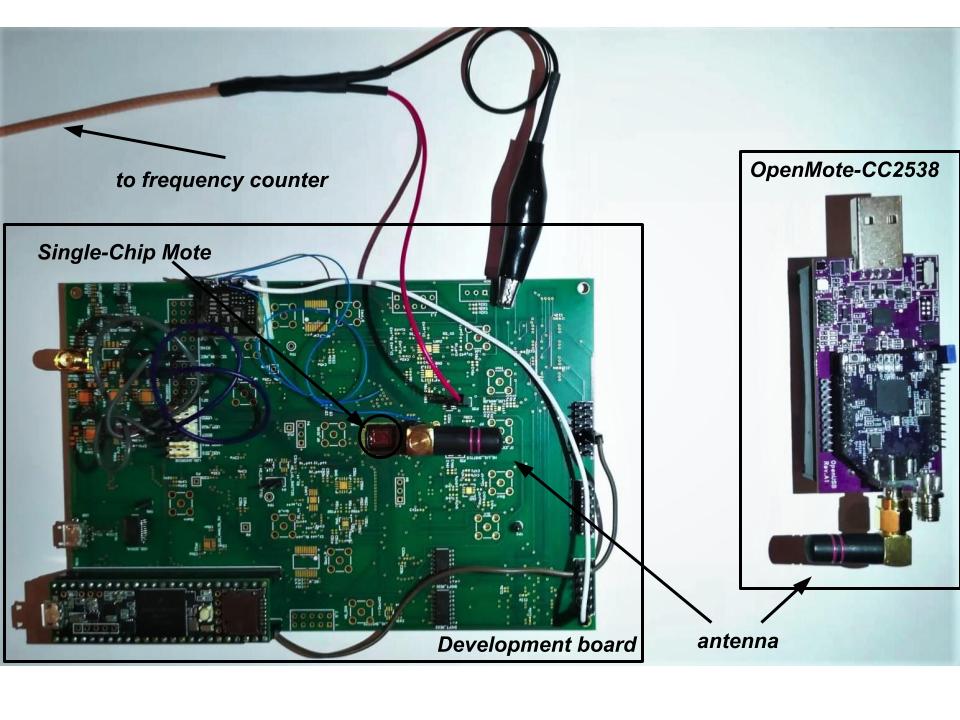}
    \caption{
        Experimental setup.
        The crystal-free single-chip mote is capable of receiving IEEE802.15.4-compliant frames from the OpenMote-CC2538 board.
    }
    \label{fig:experimental_setup_annotated}
\end{figure}


There are significant challenges to making crystal-free radios.
The main advantage of a crystal is that the frequency it oscillates at is very stable over time and temperature, typically in the 10-30~ppm range for regular crystals, even down to 2-3~ppm for temperature-
compensated versions~\cite{watteyne16crystalfree}.
On the other hand, on-chip oscillators suffer from high variations over time and temperature, which we characterize in Section~\ref{sec:characterization}. These variations would need to be detected and compensated by calibration algorithms running continuously during the lifetime of the crystal-free radio. 
Efforts have been made towards achieving an on-chip frequency reference of higher accuracy~\cite{shih13tunableOnchip,zhang11selfCal,Burn1901:CMOS}, but the temperature influence on these oscillators is still too high for compliance with IEEE802.15.4, as that standard mandates a drift below 40~ppm at all times.


The contribution of this paper is two-fold.
First, we develop an method by which a crystal-free platform with an un-calibrated on-chip oscillator tunes its radio to receive IEEE802.15.4-compliant frames, and cope with temperature variations.
The resulting algorithm tracks the IEEE802.15.4 frames it receives and continuously fine-tunes the radio to stay within the IEEE802.15.4 oscillator specifications~\cite{NXPreferenceXTAL}
The algorithm is generic and can be applied to any crystal-free platform.
Second, we implement and test the algorithm on the ``Single-Chip Mote'' (SCM)~\cite{Mesri:EECS-2016-71}, a crystal-free platform that contains a micro-controller and an ultra low-power IEEE802.15.4-compliant radio in a single chip.
We evaluate the performance of the solution by having the Single-Chip Mote communicate with an OpenMote~\cite{openmote}, a well-known crystal-based IEEE802.15.4 compliant platform.
From the best of our knowledge, this paper is the first one to show a crystal-free radio successfully communicating with a crystal-based IEEE802.15.4 compliant radio. 


The remainder of this paper is organized as follows.
Section~\ref{sec:related} surveys related work on crystal-free radios.
Section~\ref{sec:characterization} characterizes the on-chip oscillators most relevant to this paper.
Section~\ref{sec:startup} develops the algorithm for the start-up calibration of the crystal-free radio, and evaluates its performance on SCM.
Section~\ref{sec:temperature} extends the algorithm and the evaluation on maintaining calibration over temperature.
Finally, Section~\ref{sec:conclusions} concludes this paper.

\section{Related work}
\label{sec:related}


Watteyne~\etal~did some early work with the eZ430-RF2500 platform to replace the ``slow'' 32~kHz crystal by the internal oscillators of the MSP430 micro-controller on that platform~\cite{watteyne16crystalfree}. 
They implement an adaptive synchronization technique where neighbor nodes re-synchronize to one another at least every 10~s.
The resulting drift is approx. 100~ppm.
While this work does not attempt to replace the ``fast'' crystal used by the radio (which is what our paper does), it does show that getting $<40$~ppm using on-chip oscillators is a challenge.


Mehta~\etal~explore whether it is possible to relax the requirement IEEE802.15.4 puts on maximum oscillator drift requirements for the RF carrier frequency accuracy from $\pm$~40~ppm to $\pm$~1000~ppm~\cite{mehta11foffset}.
They show, by simulation, that standards-compliant narrow-band wireless communication is still feasible with $\pm$~1000~ppm oscillators by compensating their drift using
    a wide bandwidth channel-select filter,
    a demodulator and
    an adaptive feedback loop in the receiver.


Wheeler~\etal~focus on the design parameters and the electronic scheme of a crystal-free radio that has
    a drift due to phase noise less than the $\pm$~40~ppm specification of IEEE802.15.4 over 13~h in a constant-temperature environment, and
    a drift of 95~ppm$/^{\circ}C$ when the temperature changes~\cite{wheeler17}.
In order to deal with temperature variations, the authors use demodulator-based feedback to allow the receiver to track the drift of the transmitter when the transmitter is placed in a temperature chamber and subjected to a temperature variation of $2^{\circ} C/min$.
The devices used in the experiments are composed of FPGA boards and communicate over a wire on which jitter is added to emulate the wireless medium.
The receiver is able to track the transmitter's signal. 


Khan~\etal~propose a solution to calibrate the frequency of a crystal-free radio by tracking the beacons sent by a node that has a crystal reference~\cite{khan17fref}.
This work assumes the crystal-free radio is able to receive the beacons.
In order to demonstrate the feasibility of the proposal, the authors use off the shelf OpenMote-CC2538 boards~\cite{openmote}.
One OpenMote-CC2538 board is programmed to use its crystal and transmit.
The other two OpenMote-CC2538 boards simulate a crystal-free device: RX and RC connected to an FPGA that calibrates the RC oscillator.
The authors obtain an accuracy of 70~ppm for a 1~MHz RC oscillator. 


Khan~\etal~extend this work by testing their approach on a FPGA implementation of the digital system of a crystal-free mote~\cite{khan18time}.
Using a wired setting, the accuracy obtained using network calibration is 47~ppm for a 25~MHz oscillator. 


Our paper takes the state of the art one step further, as it presents a complete solution for initially tuning and maintaining the tuning so a crystal-free radio is able to receive a IEEE802.15.4-compliant frame sent by an off-the-shelf crystal-based device.
To the best of our knowledge, this is the first paper to achieve this.

\section{IEEE 802.15.4 crystal-free radio characterization}
\label{sec:characterization}

This section details the challenges is terms of clocking of a crystal-free radio.

For being able to receive and/or transmit IEEE802.15.4 frames, there are two important frequencies that need to be correctly generated by any compliant-device:
    the radio channel frequency (in the 2.4~GHz band) and
    the "chipping" frequency (2~MHz) used to modulate/demodulate the packets.
In this paper, we refer to the oscillators generating the radio channel frequency and the 2~MHz frequency as the ``RF clock'' and the ``chipping clock'', respectively.
Link-layer level time synchronization between devices that communicate over IEEE802.15.4 are out of scope of this paper, and is a well studied topic~\cite{watteyne16crystalfree}. 

When in receive mode (RX mode), the RF clock is the only one that needs calibration for setting it on the chosen communication channel.
This happens because, for RX mode, the chipping clock can be recovered from the incoming IEEE802.15.4 frames by a clock and data recovery module (CDR). 
When the platform is in transmit mode (TX mode), the on-chip oscillator generating the chipping clock has to be calibrated along with the one generating the RF clock.
The system is half duplex: at any given time, the RF clock is used either for transmit, or for receive. 


In the remainder of this section, we discuss the stability over time and temperature of the two oscillators that generate the RF clock and TX mode chipping clock, respectively.

\subsection{Stability over Time}

\begin{figure}
    \centering
    \subfloat[]{\label{fig:LC_stability}\includegraphics[width=0.9\linewidth]{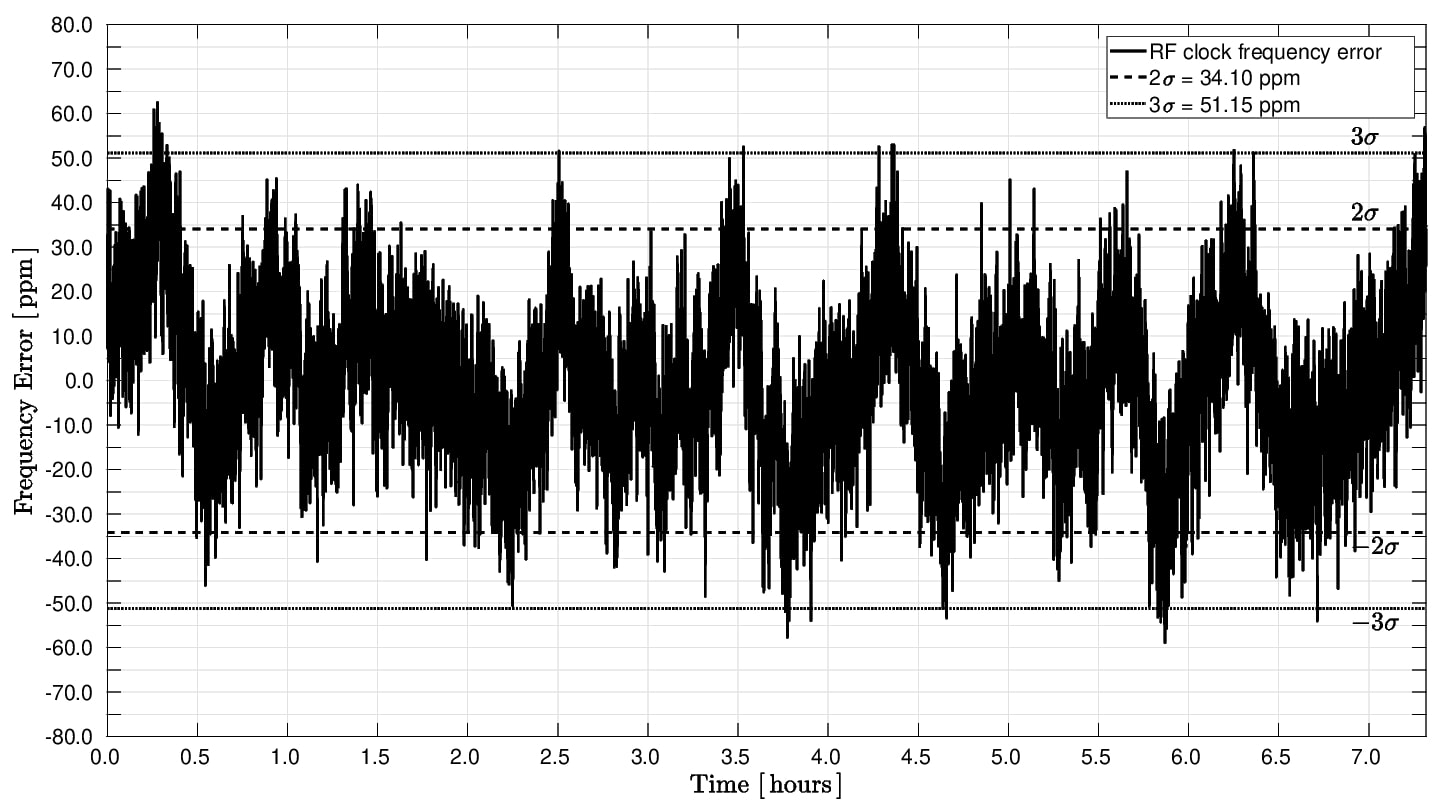}}		
    \hspace{0.25cm}
    \subfloat[]{\label{fig:M2_stability}\includegraphics[width=0.9\linewidth]{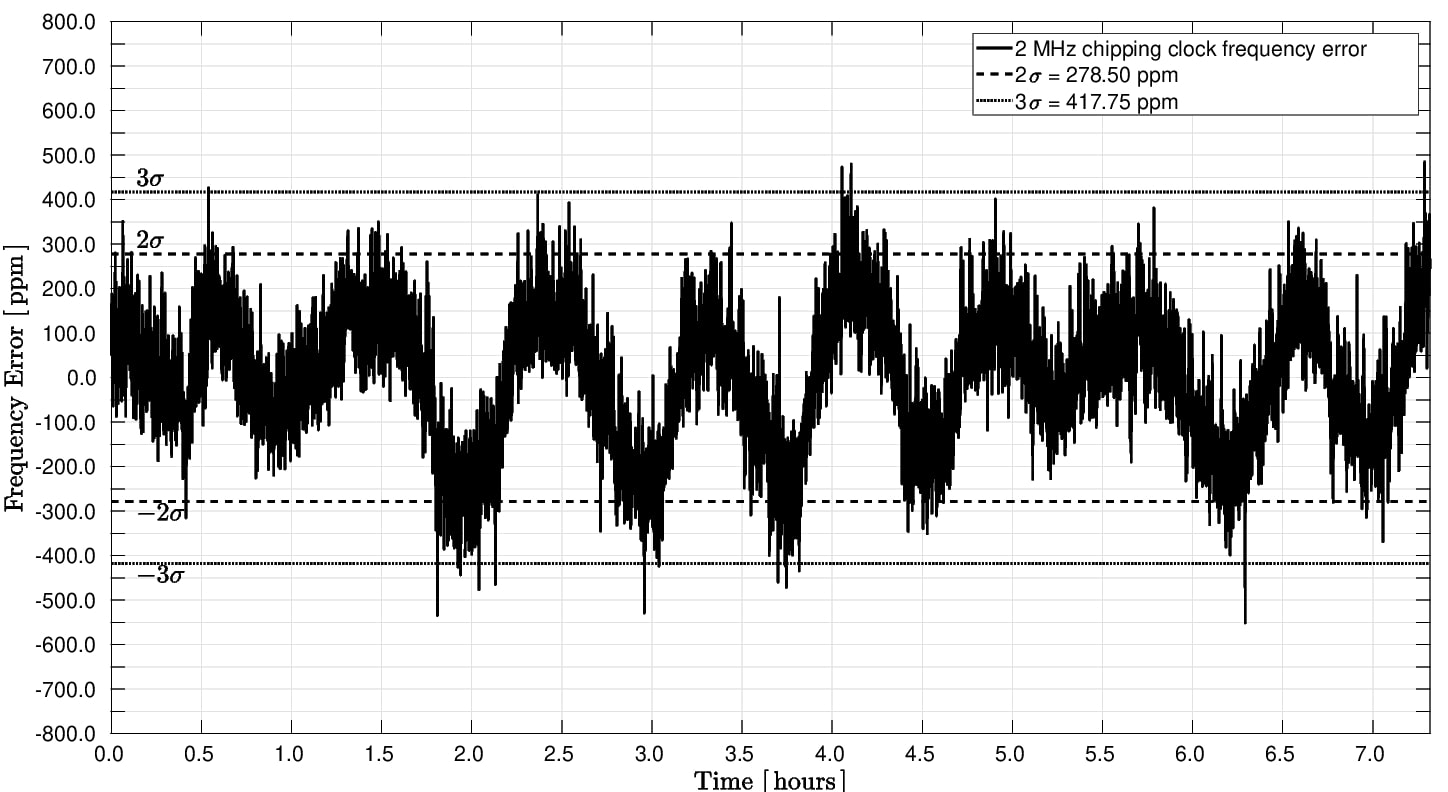}}
    \caption{
        RF (a) and chipping (b) clock frequency error during more than 7 hours of run at constant temperature ($\pm 0.3^{\circ} C$ stability).
    }
    \label{oscillators_time}
\end{figure}

An oscillator drifts over time because of temperature variations and accumulated phase noise.
To characterize the stability in time of an oscillator, the external factors that contribute to its drift need to be eliminated (in this case, the temperature variation).
Figs.~\ref{fig:LC_stability} and~\ref{fig:M2_stability} show the frequency error over a 7~h period of the RF clock and chipping clock, respectively.
For obtaining this data, the crystal-free platform was placed in a temperature chamber at constant temperature ($\pm 0.3^{\circ} C$ temperature stability) for over 7~h.
The output frequency of the two oscillators was measured using the frequency counter capability of the Agilent E4440A spectrum analyzer.

For the RF clock, in roughly 95\% of the cases ($2\sigma$), the frequency error is within $\pm$~34.1~ppm.
This means that, at constant temperature, this oscillator would meet the IEEE802.15.4 requirements of $\pm$~40~ppm accuracy. 
Some samples exceed this ($3\sigma = 51.15 ppm$), but these variations can also be a consequence of the fact that the temperature chamber used can only keep the temperature constant with a stability of $\pm 0.3^{\circ} C$.

The 2~MHz chipping clock, generated by a different oscillator, 
drifts mostly within $\pm$~278.5~ppm ($2\sigma$), when at (approximately) constant temperature.
This clock has a worse stability than the RF clock, but based on experience and measurements, the tolerable accuracy of this clock for communication can be up to $\pm$~1000~ppm, which is also documented in ~\cite{TI-CC2538}.

\subsection{Stability over Temperature}

\begin{figure}
    \centering
    \subfloat[]
    {\label{fig:LC_temp}\includegraphics[width=0.9\linewidth]{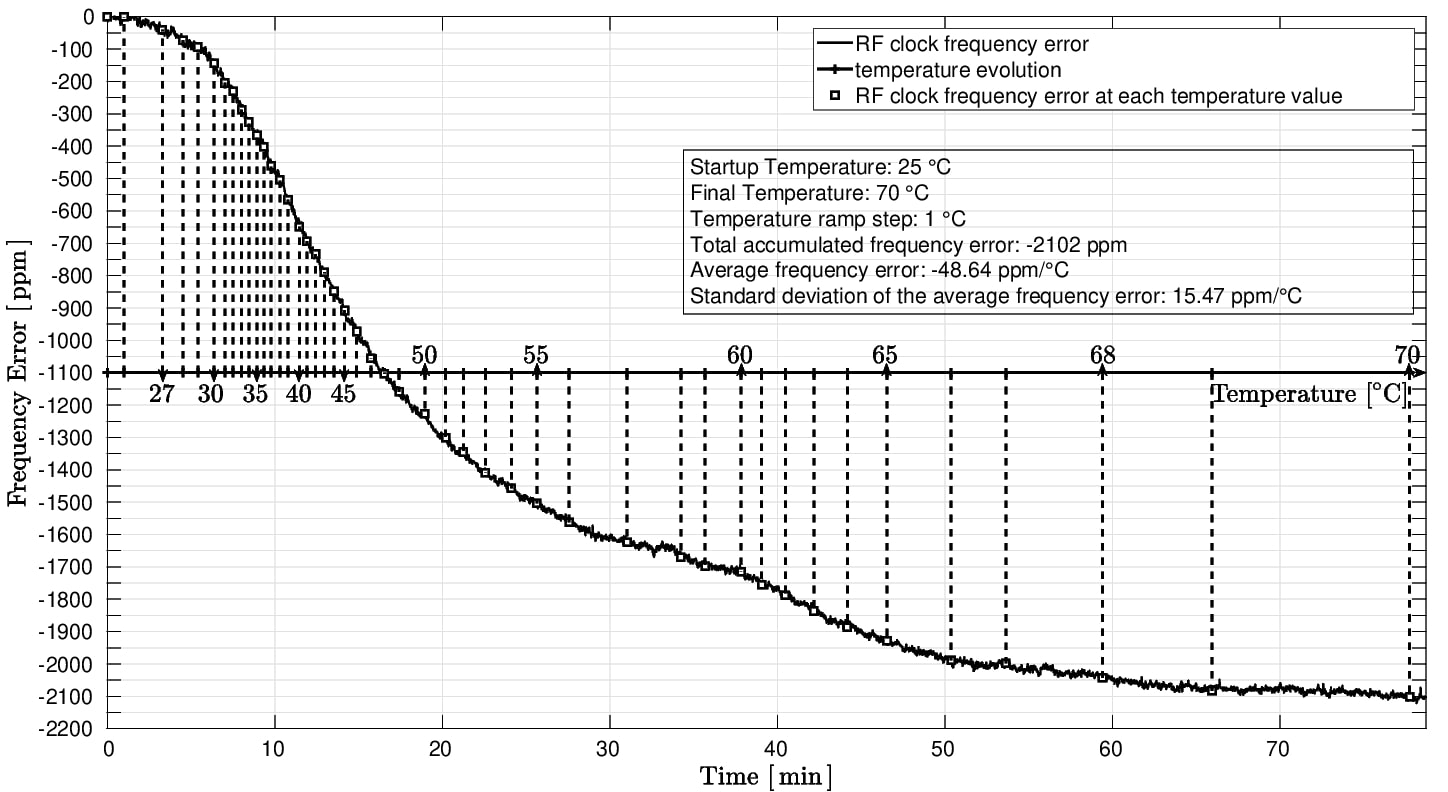}}
    \hspace{0.25cm}
	\subfloat[]
	{\label{fig:2M_temp}\includegraphics[width=0.9\linewidth]{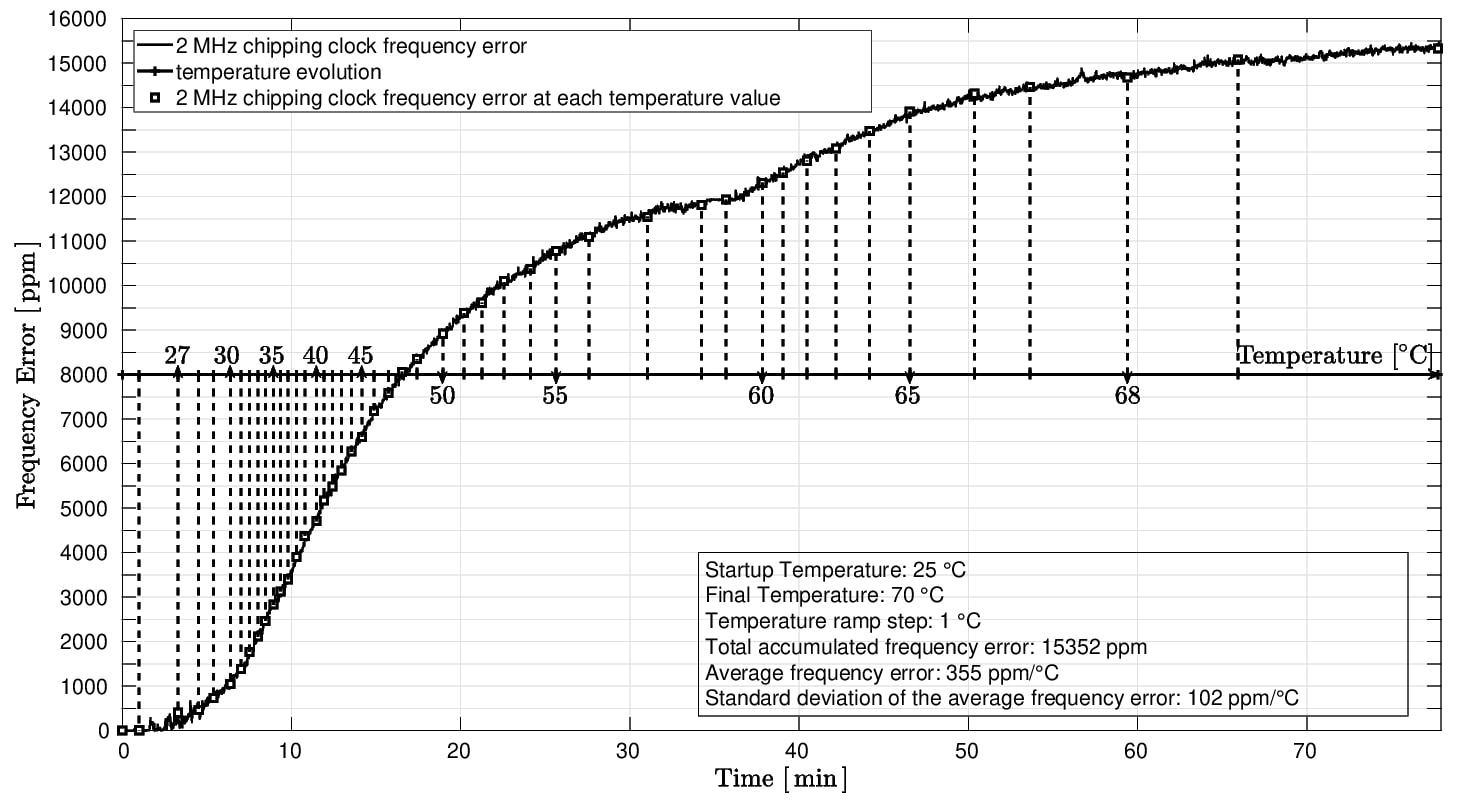}}
    \caption{\label{both_temp}RF (a) and chipping (b) clock frequency error when increasing the temperature from $25^{\circ} C$ to  $ 70^{\circ} C$.}
\end{figure}

When setting the temperature chamber to vary the temperature from $\approx 25^{\circ} C$ (room temperature) up to  $70^{\circ} C$, the two clocks experience important drifts.
Figs.~\ref{fig:LC_temp} and~\ref{fig:2M_temp} show the frequency error of the RF clock and chipping clock, respectively, as the temperature increases over time.
The two clocks behave differently.
The RF clock frequency decreases in average with $48.64 ppm/^{\circ} C$, while the chipping clock frequency increases in average with $355 ppm/^{\circ} C$.
Deviations from these values could be caused by a slightly different behavior of the oscillators at higher temperatures and/or influenced by the characteristics of the temperature chamber itself ($\pm 0.3 ^{\circ} C$ temperature stability, $\pm 3.25 ^{\circ} C$ homogeneity, $\pm 2$ \% temperature set error).

These 
significant drifts over temperature need to be corrected in order to initiate/maintain communication with an IEEE802.15.4 compliant device.

\section{Startup Calibration of a Crystal-free Radio}
\label{sec:startup}

To establish communication in a TDMA wireless network, an end device listens for a beacon sent
by
a device already in the network to get the network parameters.
It then sends a request to join the network.
After the end device and the network mutually authenticate one another, the end device learns about the communication schedule used in the network. 

When a crystal-based end device looks for beacons, it can synthesize the beacon channel frequency with a very small error, allowing it to almost immediately receive beacons and process them.
This happens because the crystals oscillate at a determined value with a very good stability ($<$10~ppm).
For a crystal-free end device, this process is more complicated and time consuming and, as there is no calibrated reference on-chip, at startup, all on-chip oscillators can be for example 10,000~ppm off their nominal value~\cite{khan17fref}.
This section describes the algorithm for 
tuning to 
the beacon channel frequency, and the method for calibrating the chipping clock. 

\subsection{Finding an IEEE802.15.4 Network}

\begin{figure}
    \centering
    \subfloat[]
    {\label{fig:SCuM_beacon_search}\includegraphics[width=0.80\columnwidth]{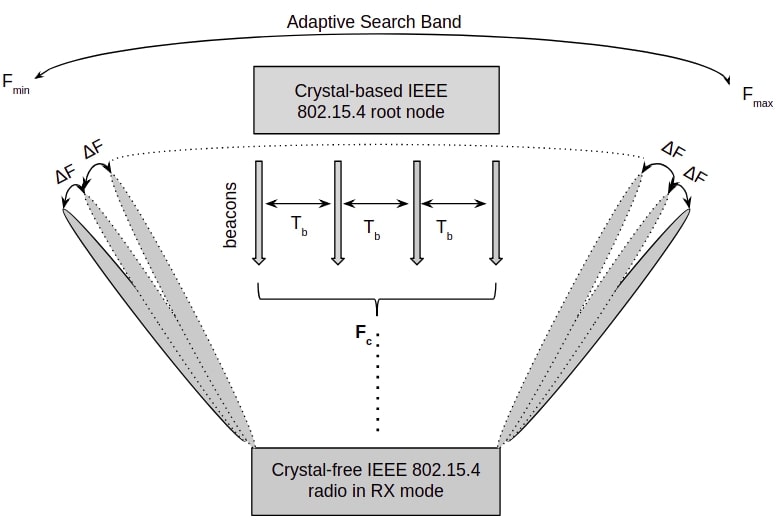}}
    \hspace{0.25cm}
    \subfloat[]
    {\label{fig:SCuM_ch_acq}\includegraphics[width=0.80\columnwidth]{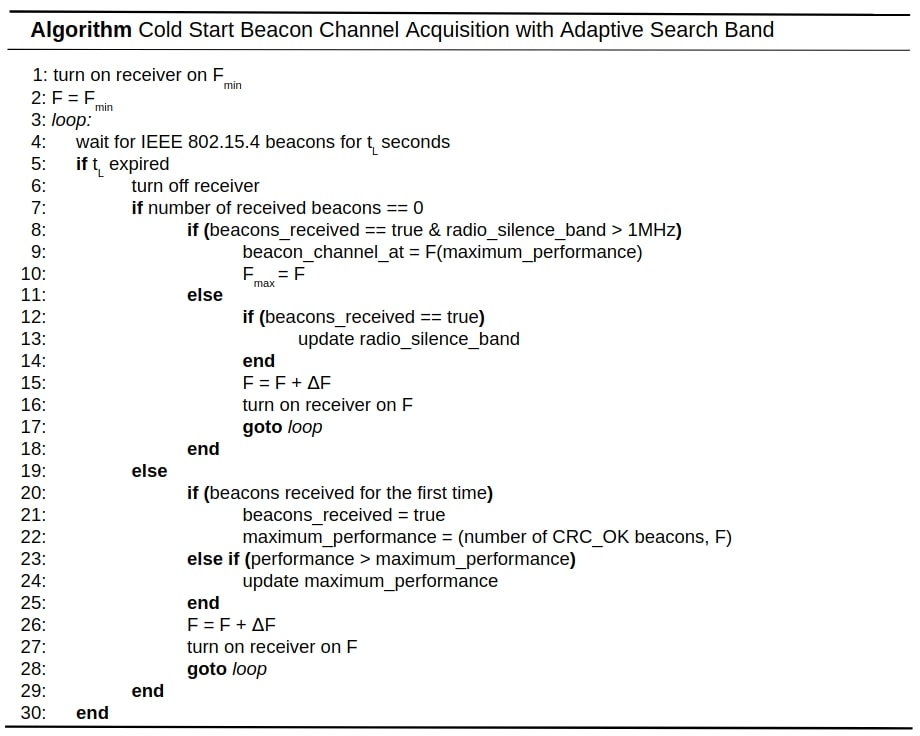}}
    \caption{
        \textit{(a)} The process of searching for the beacon channel with the center frequency $F_c$.
        \textit{(b)} Algorithm for beacon channel acquisition from cold start.
    }
\end{figure}

Fig.~\ref{fig:SCuM_beacon_search} represents the process of searching for the beacon channel frequency;
    Fig.~\ref{fig:SCuM_ch_acq} represents the corresponding algorithm.
The algorithm starts by turning the receiver on (RF clock) and listening for beacons on an unknown frequency, $F_{min}$, given by starting the RF clock on the minimum setting it supports in the 2.4~GHz band.

The reason why we qualify $F_{min}$  as ``unknown'' is two-fold:
\begin{enumerate}
    \item Fixing the RF oscillator on the minimum supported setting outputs a frequency that is highly dependent on the temperature in the environment.
    \item The resulting $F_{min}$ value varies highly from RF oscillator to RF oscillator.
\end{enumerate}

We cannot estimate $F_{min}$ so as to be able to choose a value closer to the (known) beacon channel frequency. 
Starting the algorithm at $F_{min}$ makes the algorithm board and temperature independent.
The crystal-free mote listens for $t_L$ seconds on this frequency and then updates it to $F_{min} + \Delta F$, where $\Delta F$ is the RF oscillator tuning resolution, obtained by incrementing the oscillator's minimum supported setting by 1.
In our particular case, $\Delta F \approx 90 kHz$.
The listening duration $t_L$ is chosen so as to be able to receive at least 2~beacons, based on the known interval between beacons $T_b$ (network parameter) and taking into account that the time keeping clocks are crystal-free and uncalibrated too.
For these reasons, in our particular case, $t_L = 1s$ for $T_b = 125ms$, long enough to compensate for an uncalibrated time keeping clock ($\pm$~10000~ppm).
$t_L$ could be decreased for finding the beacon channel center frequency, $F_c$, faster. 

The algorithm keeps increasing the RF clock frequency in $\Delta F$ steps, until the radio starts receiving beacons.
It then computes the maximum performance parameter and stores the setting of the RF clock at that time.
The maximum performance parameter is given by the maximum number of beacons received with a valid CRC OK.
This process continues until beacons are not received anymore (radio silence) for more than 1~MHz, to make sure that the beacon channel bandwidth (2~MHz) has been completely scanned.
At this point, the algorithm tunes the RF clock on the setting that gave the best performance, as this will be as close as possible to the center frequency of the beacon channel ($F_c$).
The beacon channel does not need to be known in advance. A timeout parameter could be set so as to restart the process in case no beacon could be received correctly.

\begin{figure}
    \centering
    \includegraphics[width=0.5\textwidth]{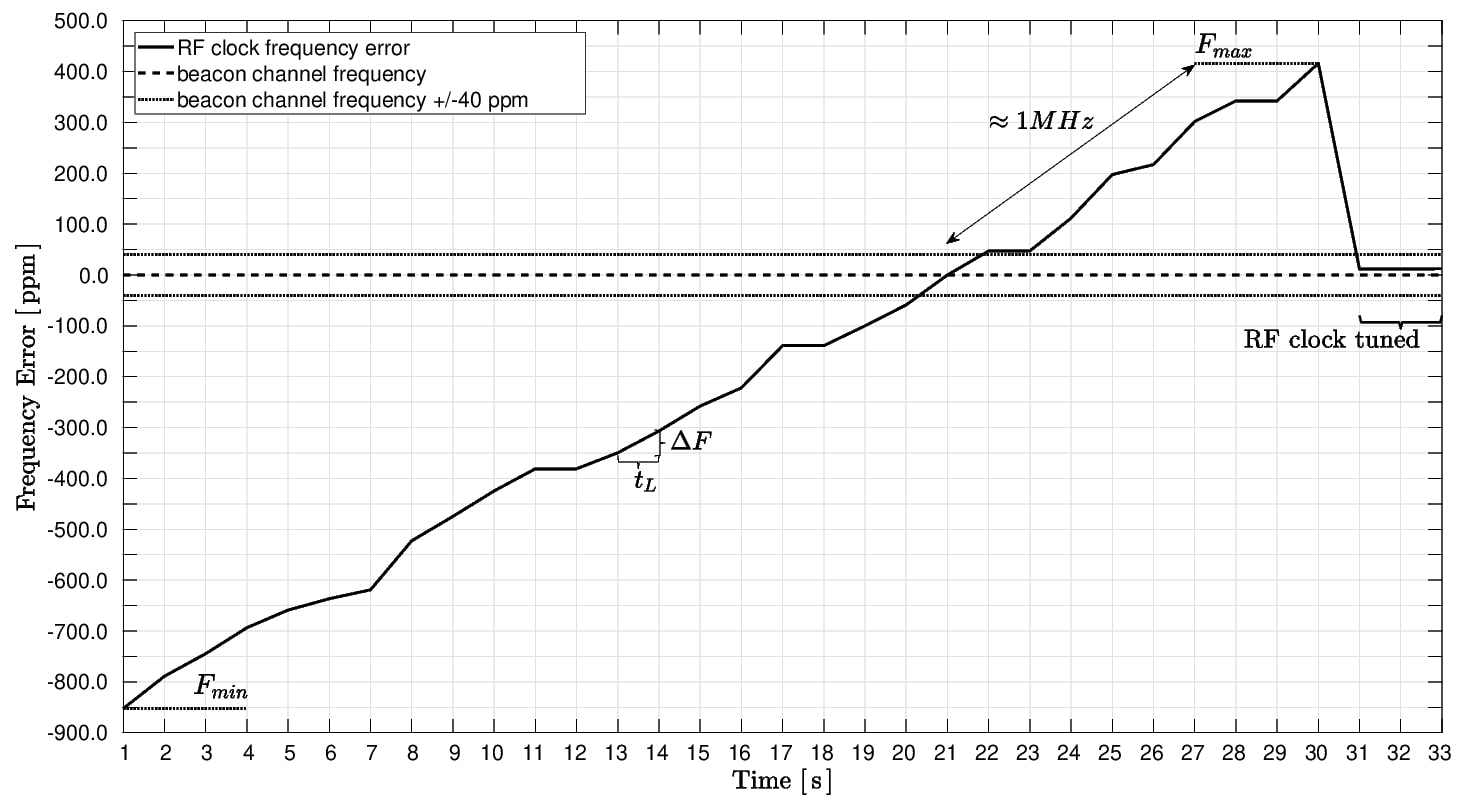}
    \caption{
        The RF clock frequency changes as the algorithm sweeps through frequencies to find that of the beacon.
        Beacons are periodically sent by an OpenMote.}
    \label{fig:jnc_fcounter}
\end{figure}

Fig.~\ref{fig:jnc_fcounter} shows the evolution of the RF clock frequency while running the beacon channel search algorithm on the crystal-free platform, measured with a frequency counter.
We can see that, at startup, the RF clock is -850~ppm away from the center frequency of the beacon channel.
The RF frequency is increased gradually, and after 1~MHz of radio silence the frequency sweep ends and the RF clock is tuned to the setting that yielded the best performance.
The resulting RF frequency is within $\pm$~40~ppm of the beacon center frequency.
Beacons are sent every 125~ms ($T_B$) by an OpenMote and the listening duration ($t_L$) is $\approx$~1~s.

\subsection{Calibrating the Chipping Clock}
\label{sec:cal_2M}

Once the RF clock is tuned to the center frequency of the beacon, the crystal-free platform can receive the beacons sent every $T_B$ by the OpenMote.
To calibrate the chipping clock, one needs to compare the number of ticks between two successive beacons with the value an ideal 2~MHz clock would count, and apply a correction based on its tuning resolution ($\Delta F_{2MHz}$):
$$correction = -\frac{ticks_{chipping}-ticks_{ideal2MHz}}{\Delta F_{2MHz}}$$
We call this correction ``fast calibration,'' as it allows one to quickly bring the chipping clock closer to the 2~MHz ideal value.
After this is done, more fine corrections can be applied to the chipping clock by adjusting its oscillator setting with $\pm$~1.
These finer corrections can be applied at each received beacon, or after the average number of ticks counted during the last $N$ received beacons is compared to a threshold value (a calibration window defining the accepted error of the average counted ticks of $\pm x ppm$).
This latter method is discussed in~\cite{khan18time} and proves to have better accuracy than the former.

\begin{figure}
    \centering
    \includegraphics[width=0.5\textwidth]{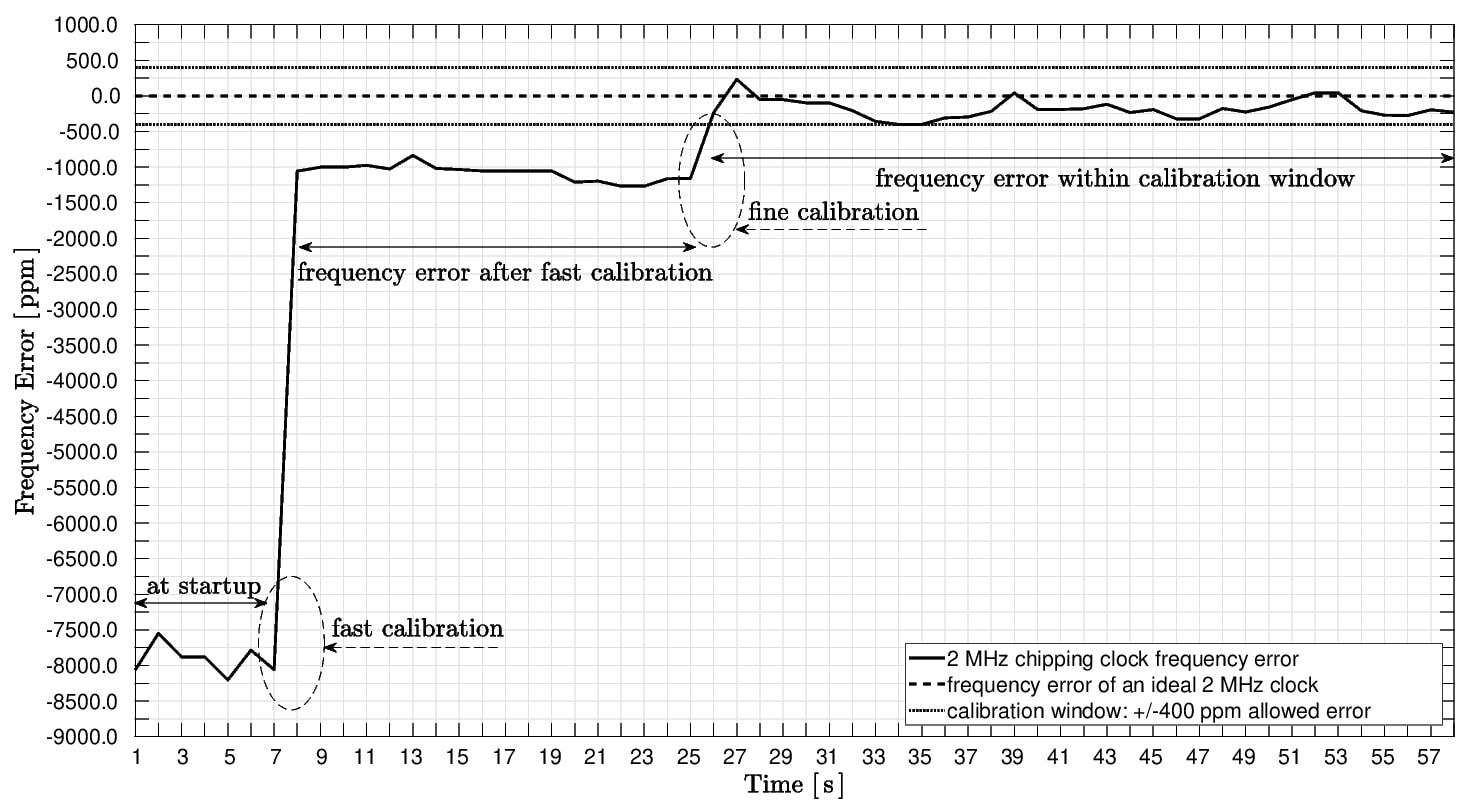}
    \caption{
        2~MHz chipping clock frequency evolution while running the fast calibration strategy followed by fine clock corrections, laboratory environment.
        Beacons are periodically sent by an OpenMote.
    }
    \label{fig:M2_fastcal}
\end{figure}

We implement the fast calibration strategy on the crystal-free platform, as well as the fine corrections based on the average number of ticks counted by the chipping clock.
We set $N = 10$ received beacons before calibration, and a calibration window of $\pm x = \pm 400 ppm$ accepted frequency error.
We extract data with a frequency counter, in laboratory environment.
Fig.~\ref{fig:M2_fastcal} shows how at startup (while the RF clock looks for the beacon channel), the chipping clock is 8000~ppm away from the nominal 2~MHz value.
After fast calibration, it gets within 1000~ppm of the nominal value.
After fine calibrations, the chipping clock reaches a frequency error variation inside the defined calibration window. 
The interval of time between fast calibration and fine calibration is influenced by $N$ and by the number of lost beacons: the more beacons are lost, the slower the algorithm reaches the $N$ needed packets before the next calibration round. 

\section{Maintaining Calibration over Temperature}
\label{sec:temperature}

The methods presented in Section~\ref{sec:startup} can be applied to calibrate the two clocks of the crystal-free mote at startup, no matter the temperature of the environment.
After startup, the frequency errors experienced by the RF and chipping clocks are strongly influenced by temperature variation.
This is shown in Fig.~\ref{both_temp}.
This section presents a method for staying calibrated after startup even when the temperature changes. 

\subsection{Correcting the RF clock}

As in any standard receive chain, when the RF clock is tuned to the beacon frequency and the crystal-free mote receives a beacon, the received RF signal is down-converted to a specified intermediate frequency (IF) and then demodulated~\cite{mehta11foffset}.
The IF frequency is a fixed hardware design parameter and can be measured during packet reception.
As the temperature in the environment changes, the RF frequency drifts ($\approx 48.64 ppm/^{\circ}C$) and the measured IF frequency shows an offset from the expected value, when receiving a packet. 

\begin{figure}
    \centering
    \includegraphics[width=\columnwidth]{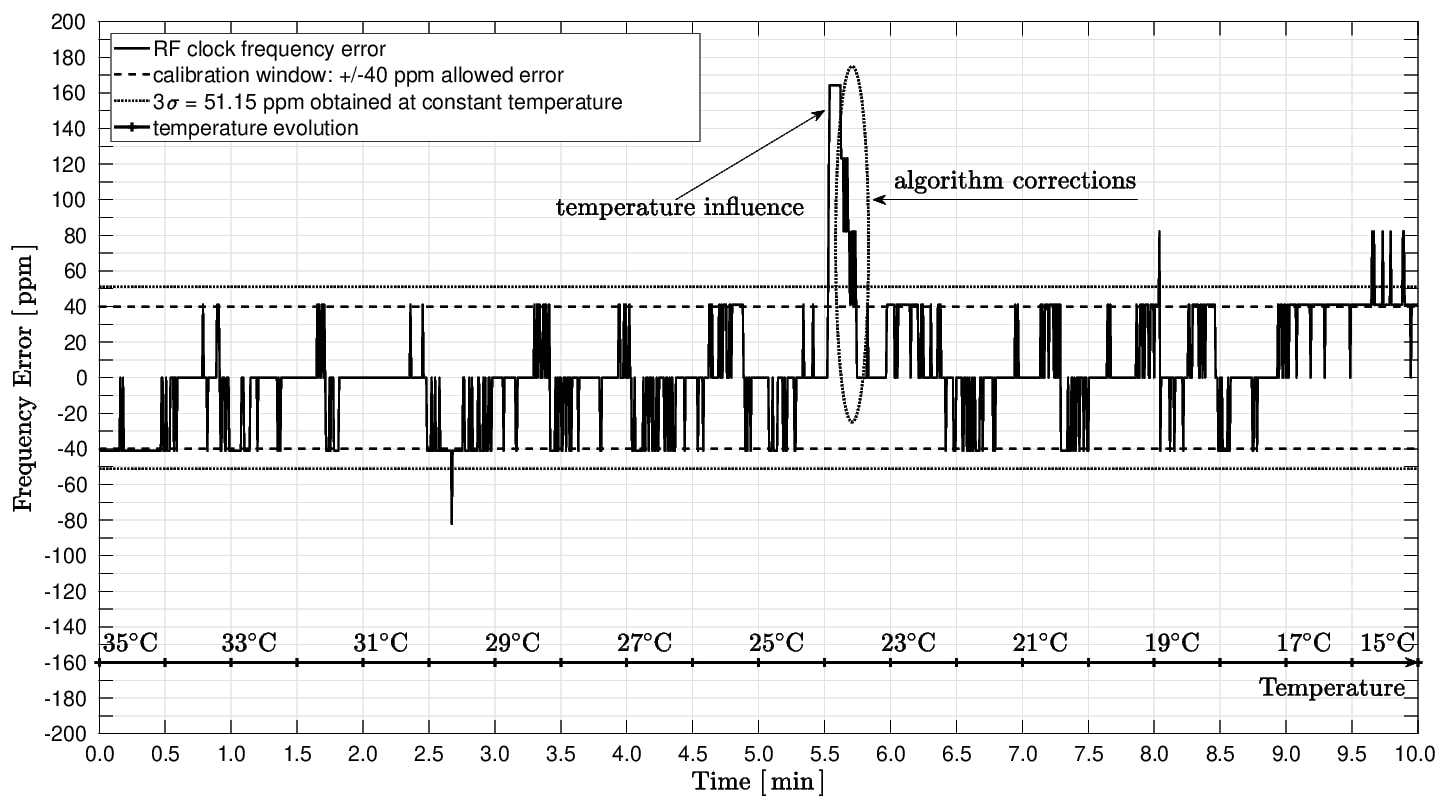}		
    \caption{
        RF clock corrections when subjected to a $2^{\circ}C/min$ temperature variation.
        The RF clock is kept within $\pm$~40~ppm.
        Beacons are sent periodically by an OpenMote.
    }
    \label{fig:IFcal_LC}
\end{figure}

We implement an algorithm that checks the offset of the IF frequency at each beacon reception, and use that to finely tune the RF clock ($\pm \Delta F$, depending on the sign of the offset), as this clock is already calibrated on the beacon frequency and only small corrections are needed.
This calibration comes at no cost, as the IF information is available at every packet reception.
Fig.~\ref{fig:IFcal_LC} shows the evolution of the RF clock frequency error while the crystal-free mote is placed inside a temperature chamber and is subject to a $15^{\circ} C$ temperature change.
During this time, the crystal-free platform finely tunes the RF clock based on the measured IF every time a beacon from the OpenMote is received ($T_B = 125ms$).
The obtained RF frequency error is as good as in constant temperature environment.
The frequency error spike in Fig.~\ref{fig:IFcal_LC} is a consequence of several lost beacons, but it is corrected as soon as new beacons are received. This is possible because, even if the receiver's RF clock may have accumulated more than 40~ppm of error, the beacons are still within the bandwidth of the receiver.

The algorithm is packet-loss tolerant, as long as the temperature variation isn't too large and causes the RF clock to drift outside the communication channel between two packet receptions.

\subsection{Correcting the Chipping Clock}

As long as the RF clock is continuously corrected and the crystal-free platform is able to receive beacons against temperature changes in the environment, the 2~MHz chipping clock errors due to temperature can be corrected using the fine calibration method described in Section~\ref{sec:cal_2M}.

\begin{figure}
    \centering
    \includegraphics[width=0.5\textwidth]{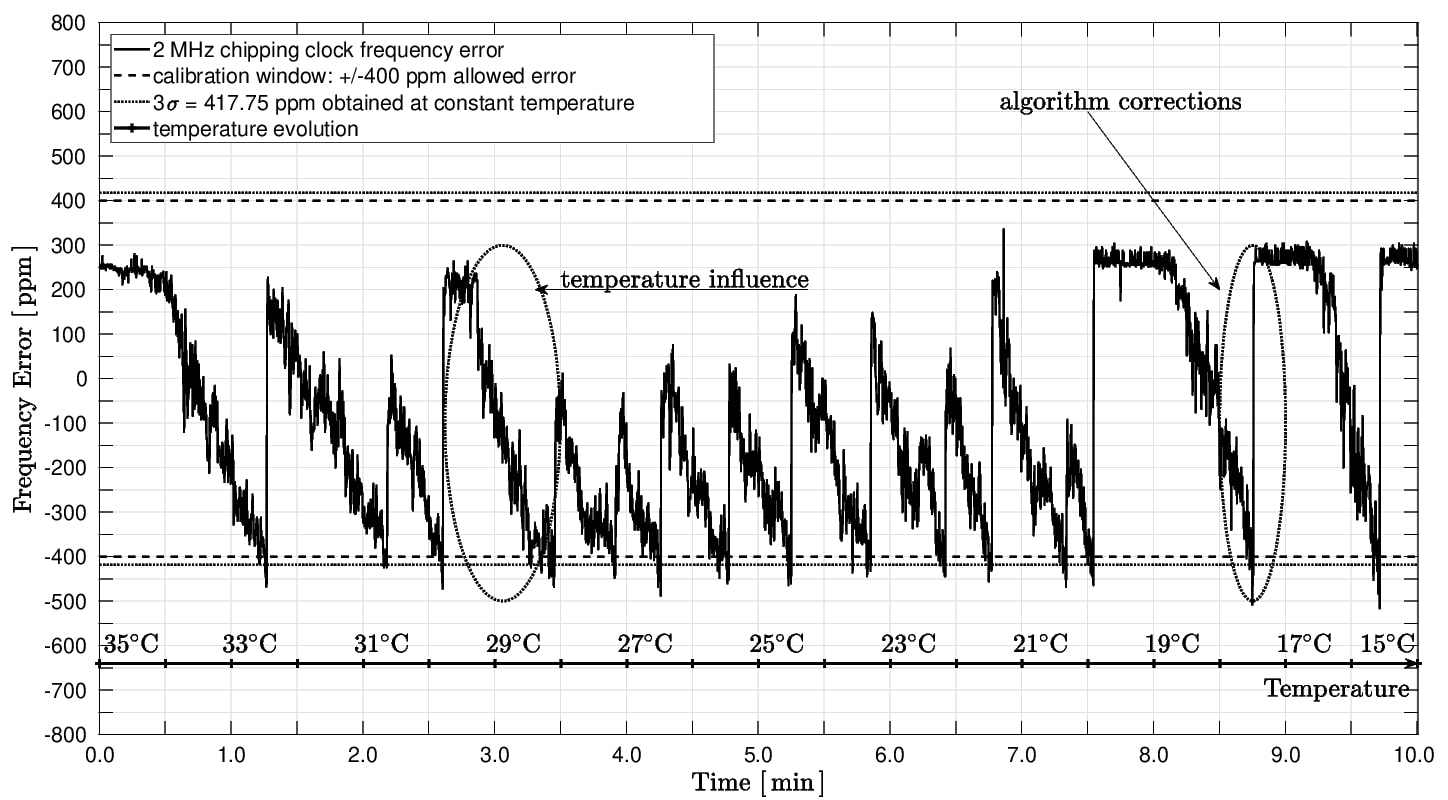}
    \caption{2 MHz chipping clock corrections when subjected to a $2^{\circ}C/min$ temperature variation. The clock is kept within the $\pm 400 ppm$ calibration window. Beacons are periodically sent by an OpenMote.}
    \label{fig:IFcal_2M}
\end{figure}

Fig.~\ref{fig:IFcal_2M} shows the performance of the applied fine calibration: the frequency error of the 2~MHz chipping clock is as good as at constant temperature, even when the crystal-free platform is subjected to a $15^{\circ} C$ temperature variation.

\section{Conclusions}
\label{sec:conclusions}

In the context of crystal-free radios, this paper analyzes the frequency stability with respect to time and temperature of two fundamental on-chip oscillators.
These oscillators drive the wireless communication capabilities of a radio in typical IEEE802.15.4 networks.
While the frequency stability in time is good enough to meet the specification demands, these oscillators experience very 
significant
drift over temperature.

We present a mechanism to dynamically compensate that drift to first enable the device to find the appropriate frequency to join the IEEE802.15.4 network, and then to calibrate other on-chip oscillators to support for example chirp decoding.
We show that, using the IEEE802.15.4 signaling, we can keep the obtained calibration even as the temperature changes. 
The obtained accuracy meets the IEEE802.15.4 requirements of $\pm 40 ppm$ frequency stability.
All presented strategies are accompanied by experimental results obtained with a crystal-free platform and a crystal-based standards-compliant OpenMote acting as join proxy node of the IEEE802.15.4 network. 

\section*{Acknowledgment}
This work has received funding from the European Union Horizon 2020 research and innovation programme under the Marie Sklodowska-Curie grant agreement No 675891 (SCAVENGE project) and was conducted with the support of the Berkeley Sensor and Actuator Center (BSAC) and the Berkeley Wireless Research Center (BWRC). This work is also partially supported by the Spanish Ministry of Economy and the ERDF regional development fund under SINERGIA project (TEC2015-71303-R). 

\bibliographystyle{IEEEtran}
\bibliography{suciu19experimental}

\end{document}